\documentclass{aa}  

\usepackage{graphicx}
\usepackage{txfonts}
\usepackage[colorlinks=true,citecolor=blue]{hyperref}
\usepackage{soul}

\titlerunning{Galaxy Clusters with SIDM}
\authorrunning{Ragagnin et al.}

\begin{document} 


   \title{Dianoga SIDM: galaxy cluster self-interacting dark matter simulations
   }

 \author{Antonio Ragagnin,\inst{\ref{oas},\ref{unibo},\ref{ifpu}}\thanks{\email{antonio.ragagnin@unibo.it}}
   M. Meneghetti,\inst{\ref{oas},\ref{infnbo}}
   F. Calura,\inst{\ref{oas}}
   G. Despali,\inst{\ref{unibo},\ref{oas},\ref{infnbo}}
   K. Dolag,\inst{\ref{usm},\ref{mpia}}
   M. S. Fischer,\inst{\ref{usm},\ref{origins}}
   C. Giocoli,\inst{\ref{oas},\ref{infnbo}}
   L. Moscardini\inst{\ref{unibo},\ref{oas},\ref{infnbo}}
   }
   \institute{
    INAF-Osservatorio di Astrofisica e Scienza dello Spazio di Bologna, Via Piero Gobetti 93/3, I-40129 Bologna, Italy\label{oas}   \and
    Dipartimento di Fisica e Astronomia "Augusto Righi", Alma Mater Studiorum Università di Bologna, via Gobetti 93/2, I-40129 Bologna, Italy\label{unibo} \and
            IFPU - Institute for Fundamental Physics of the Universe, Via Beirut 2, I-34014 Trieste, Italy\label{ifpu} \and
    INFN-Sezione di Bologna, Viale Berti Pichat 6/2, I-40127 Bologna, Italy\label{infnbo}
            \and Universit\"ats-Sternwarte, Fakult\"at für Physik, Ludwig-Maximilians-Universit\"at M\"unchen, Scheinerstr. 1, D-81679 M\"unchen, Germany\label{usm}
            \and Max-Planck-Institut f\"ur Astrophysik, Karl-Schwarzschild-Str. 1, D-85748 Garching, Germany\label{mpia}
            \and Excellence Cluster ORIGINS, Boltzmannstrasse 2, D-85748 Garching, Germany\label{origins}
            }

   \date{submitted}

 
  \abstract
   {Self-interacting dark matter (SIDM) can tackle or alleviate small-scale issues within the cosmological standard model $\Lambda$CDM,  and diverse flavours of SIDM can produce unique astrophysical predictions, resulting in different possible signatures which can be used to test these models {with dedicated observations of galaxy clusters}.}
   {This work  aims at assessing the impact of DM self-interactions on the properties of galaxy clusters. In particular, the goal is to study the angular dependence of the cross section by testing rare (large angle scattering) and frequent (small angle scattering) SIDM models with velocity-dependent cross sections.}
   {We re-simulate six galaxy cluster zoom-in initial conditions with a dark matter only run and with a full-physics setup simulations that includes a self-consistent treatment of baryon physics. We test the dark matter only setup and the full physics setup  with either collisionless cold dark matter, rare self-interacting dark matter, and frequent  self-interacting dark matter models. We then study their matter density profiles as well as their subhalo population.}
   {Our dark matter only SIDM simlations agree with theoretical models, and when baryons are included in simulations, our SIDM models substantially increase the central density of galaxy cluster cores compared to full-physics simulations using collisionless dark matter. SIDM subhalo suppression in full-physics simulations is milder compared to the one found in dark matter only simulations, because of the cuspier baryionic potential that prevent subhalo disruption. Moreover SIDM with small-angle scattering significantly suppress a larger number of subhaloes compared to large angle scattering SIDM models. Additionally, SIDM models generate a broader range of subhalo concentration values, including a tail of more diffuse subhaloes in the outskirts of galaxy clusters and a population of more compact subhaloes in the cluster cores.
   }
   {}


   \maketitle
%

\section{Introduction}

Galaxy clusters, recognised as the most extensive gravitationally bound systems of galaxies~\citep[see][for a comprehensive review]{2012ARA&A..50..353Kravtsov}, play a pivotal role as cosmic laboratories for investigating the large-scale structures within our Universe.  
The underlying cosmological model 
impacts their masses~\citep{Tinker2008,2000PhRvL..84.3760Spergel,2016MNRAS.456.2486Despali,2017MNRAS.469.1997Despali}, 
their lensing signal~\citep{PNKneib1997,1998ApJ...499L...5Moore}, the abundance of substructures within them~\citep{1998MNRAS.299..728Tormen,PNSpringel2004}, and their 
concentration~\citep{giocoli12b,giocoli13,Ragagnin2021MCcosmi}.  

The current standard concordance cosmological model $\Lambda$CDM assumes the presence of a cosmological constant  and a type of dark matter (DM) that is cold and collisionless. 
Numerical simulations performed within this paradigm are currently showing some tensions with observational data.
For instance simulations of galaxy cluster cores~\citep[as in][]{2020Sci...369.1347Meneghetti} reveal subhaloes that are considerably less compact than their observed counterparts~\citep[as illustrated in][]{2019A&A...631A.130Bergamini,2022A&A...659A..24Granata}.

Specifically, the galaxy-galaxy strong lensing (GGSL) signal resulting from simulated substructures of mass $M\approx 10^{11}{\rm M}_\odot$ can be up to a factor of $2$ lower than observed values~\citep{2022A&A...668A.188Meneghetti,2022A&A...665A..16Ragagnin,Meneghetti2023PersistentExcess}.
Concerning the higher mass range of satellites ($M>10^{11}\,{\rm M}_\odot$), hydrodynamic simulations can indeed reproduce high mass subhaloes with compactness as high as the one from scaling relations~\citep[see the simulations presented in][]{2021MNRAS.505.1458Bahe,2021MNRAS.504L...7Robertson}. This can be achieved, for instance, by assuming particularly low AGN efficiencies. However, it is noteworthy that the lensing signal derived from observational data is predominantly influenced by satellites with lower masses~\citep{2022A&A...665A..16Ragagnin}.
In fact the study conducted by \cite{2022A&A...665A..16Ragagnin} indicates that relying solely on AGN physics does not appear to find a solution to the compactness mismatch without compromising the consistency with observed properties related to stellar mass.

Furthermore, $\Lambda$CDM simulations face challenges in reproducing ultra-diffuse galaxies~\citep[UDGs, as studied by][]{VanDokkum2015ComaUDGs,Mowla2017UDG,Greco2018UDG}, whose origin and low concentration parameter may be explained by satellites with cored dark matter profiles \citep{Carleton2019UDGsFromCored}.
Although baryonic processes can also play a crucial role in lowering subhalo compactness~\citep{DiCintio019UDGs}, $\Lambda$CDM simulations still fall short in reproducing the UDGs low circular velocity \citep[see][particularly their Fig. 5]{Sales+2020UDGplot}.

In this paper, we explore the impact of a type of dark matter that is not collisionless. In particular, we study self-interacting  DM  \citep[SIDM,][]{2000PhRvL..84.3760Spergel} in the context of hydrodynamical simulations of galaxy clusters. Motivated by particle physics, SIDM models consider interactions between DM elementary particles and a massive mediator through, for instance, a Yukawa potential \citep{Loeb2011DwarfsAndUDGs}. SIDM maintains the theoretical expectations of cold and collisionless DM   at large scales while significantly influencing subhalo properties \citep[see, for example,][]{2000PhRvL..84.3760Spergel,Tulin2018SIDMreview,Adhikari2018subSIDMTests,MastromarinoDespali2023}, resulting in objects with a broader range of concentration parameter compared to collisionless CDM.

Notably, DM self-interactions smooth out the density distribution in the central regions of DM subhaloes, leading to profiles that are more cored than their collisionless counterparts lowering the subhalo compactness~\citep[see, for instance,][]{Carleton2019UDGsFromCored,Nadler2023arXivUDG,Kong2022UDGlow-c}.
Moreover SIDM subhaloes that formed with an initially high concentration, will eventually reveal, after the core expansion reaches its minimum, a core-collapse stage that will make them much more compact than their analogues in the $\Lambda$CDM model~\citep{Outmezguine202corecollapse,Zeng2023SubhaloCoreCollapse}, thus explaining the excess of subhalo compactness observed in galaxy clusters~\citep{2021PhRvD.104j3031Yang}.

In this work, we compare two classes of SIDM models. The first is the so-called frequent SIDM (fSIDM) model, which assumes small-angle scattering, in the limit of keeping the momentum transfer cross section fixed but decreasing the scattering angles. The second is the so-called rare SIDM (rSIDM) model, which assumes isotropic scattering. Under these conditions,  scattering events are much more frequent in the fSIDM than in the rSIDM model, hence the adjectives frequent and rare.
Simulations with a self-consistent treatment of baryonic physics (hereafter FP, as full physics) are necessary to properly study and capture the properly the evolution of galaxy cluster substructures~\citep{Kimmig2023}. Therefore in this work we plan to re-simulate our galaxy clusters both with dark matter only (DMO) simulations and FP ones.

The two models can alter the DM distribution in different ways, with fSIDM being more efficient in generating DM-galaxy offsets during mergers because of its small-angle dependency~\citep{2021MNRAS.505..851Fischer}. 
We use the implementation presented in ~\cite{2021MNRAS.505..851Fischer} and ~\cite{2022MNRAS.516.1923Fischer},  a technique that relies on describing the self-interaction between DM particles through
an effective drag force, as presented in ~\cite{Kahlhoefer_2014}.

The studies of ~\cite{Sagunski2021}, ~\cite{Andrade2022},~\cite{Harvey2019}, and \cite{2022A&A...666A..41Eckert} established a lower limit on the impact of SIDM on cluster scales. Specifically, considering that clusters adhere to an Einasto profile~\citep{1965TrAlm...5...87Einasto}, the work of 
\cite{2022A&A...666A..41Eckert} shows that the total cross section per unit dark matter particle mass $m_{\chi}$ should be constrained to $\sigma/m_{\chi}<0.19\,{\rm g}^{-1}\,{\rm cm}^2$. This value is notably low when compared to the typical values found in theoretical studies, which hover around $\sigma/m_{\chi} \approx 1\,{\rm g}^{-1}\,{\rm cm}^2$~as used in \cite{MastromarinoDespali2023} (note however that they studied a mass-range that is different than ours).

While this might initially appear as a challenge for SIDM, it is important to note that elementary particle interactions, such as those governed by the Yukawa potential, exhibit cross sections which depend on the relative velocity of the particles. The cross section decreases as a function of the relative velocity, making it applicable for constraining both dwarf spheroidal profiles and group cores~\citep{Correa2021vdSIDM}. Consequently, in this study, our focus will be solely on SIDM models with a velocity-dependent cross section.


The paper is structured as follows: we present our suite of zoom-in simulations with collisionless DM and SIDM in Sect. \ref{sec:numeric};
we investigate the overall DM distribution in the cluster cores in Sect. \ref{sec:profi}; we study the cluster subhalo population in Sect. \ref{sec:sh}; we draw our conclusions in Sect. \ref{sec:conclu}. 
Throughout this paper, we use the term "collisionless" to describe the cold and collisionless DM assumed in the standard $\Lambda$CDM, distinguishing it from SIDM, which is also a type of cold DM.

\section{Numerical Setup}
\label{sec:numeric}

\begin{table}
\caption{Regions re-simulated in this work. Different columns from left to right report the name, the virial mass, and the NFW concentration at $z=0.2.$ We present the concentration parameterson the third and fourth columns, computed both in  the DMO and FP runs respectively.} 
\label{tbl:sims}  
\centering                                      
\begin{tabular}{c c c c}     
\hline\hline                        
region & $M_{\rm vir}[10^{14}h^{-1}{\rm M}_\odot]$ & $c_{\rm vir}$ DMO & $c_{\rm vir}$ FP\\ 
\hline
D3 & $4.8$ & $4.0$ & $4.6$ \\
D4 & $2.7$ & $4.6$ & $3.9$ \\
D5 & $1.2$ & $6.0$ & $5.5$ \\
D10 & $11.2$ & $4.1$ & $4.5$ \\
D15 & $11.5$ & $4.8$ & $5.9$ \\
D16 & $12.3$ & $4.6$ & $5.0$ \\
\hline
\end{tabular}
\tablefoot{
We show mass and concentration of collisionless DM only, because these value are almost identical for the SIDM runs.}
\end{table}

Our simulations are performed using OpenGadget3,
a code derived from P-Gadget3, a successor of P-Gadget2 \citep{2005MNRAS.364.1105Springel}. The initial conditions mirror those employed in the Dianoga simulations~\citep[as utilised, for instance, in][]{2011MNRAS.418.2234Bonafede,2015ApJ...813L..17Rasia}. These conditions are generated from a parent DMO box with a side length of $1$ comoving $h^{-1}\,{\rm Gpc}$ and are specifically tailored for conducting hydrodynamic simulations of galaxy clusters.
The power spectrum assumed in the parent box corresponds to a $ \Lambda$CDM model with cosmological parameters $\Omega_m = 0.24, \Omega_b = 0.04, n_s = 0.96, \sigma_8 = 0.8,$ and $h = 0.72.$\footnote{Note that SIDM is a type of collisional, cold DM, therefore the three DM models (collisionless DM, rSIDM, fSIDM) can consistently share the same initial conditions (as opposed to what would happen with warm dark matter).}
In this work,  we re-simulated the initial conditions for six galaxy cluster regions that encompass  virial masses in the range $10^{14}-10^{15}{\rm M}_\odot$  and analyse the results in $15$ redshift slices between $z=0.2-0.6.$\footnote{ 
Note that the virial mass $M_{\rm vir}$ is defined as the mass  within the so-called virial radius $R_{\rm vir},$  that encloses an average density of $4/3\pi R_{\rm vir}^3 \Delta_{\rm vir}\rho_c,$ where $\rho_c$ is the critical density of the Universe and $\Delta_{\rm vir}$ is the overdensity coming from the top-hat spherical collapse model~\citep[see][for a review]{2018AstL...44....8Kravtsov} and has a value of $\Delta_{\rm vir}\approx100$ for our cosmological parameters \citep{eke96,1998ApJ...495...80Bryan}.
}

We maintain the same resolution level as presented in \cite{2022A&A...665A..16Ragagnin}. Specifically our simulations employ a gravitational softening of $\epsilon_{\rm DM} = 3.7\, h^{-1}{\rm ckpc}$ for DM particles and $\epsilon_\star = 2.0\,h^{-1}{\rm ckpc}$ as the softening parameter for stellar gravitational interactions. The DM particle masses are set to $m_{\rm DM} = 8.3\times10^8h^{-1} {\rm M}_\odot.$

\subsection{Baryon physics}

To follow the baryon physics, we simulate the hydrodynamics of gas using an enhanced smoothed particle hydrodynamics (SPH) solver presented in \cite{2016MNRAS.455.2110Beck}.\footnote{We used a space-filling curve-aware neighbour search~\citep{2016pcre.conf..411Ragagnin}.}
The stellar evolution scheme from~\cite{2007MNRAS.382.1050Tornatore} which follows  $11$ chemical elements (H, He, C, N, O, Ne, Mg, Si, S, Ca, Fe) with input cooling tables generated using the CLOUDY photo-ionisation code \citep{1998PASP..110..761Ferland}.
For a comprehensive understanding of the modelling of supermassive black holes and energy feedback, readers can refer to \cite{2005MNRAS.361..776Springel,2010MNRAS.401.1670Fabjan} and \cite{2014MNRAS.442.2304Hirschmann}, where prescriptions for black hole growth and feedback from AGNs are thoroughly described. The identification of haloes and their member galaxies is accomplished using the friends-of-friends halo finder~\citep{1985ApJ...292..371Davis} and an improved version of the subhalo finder SUBFIND \citep{2001MNRAS.328..726Springel}, which accounts for the presence of baryons \citep{2009MNRAS.399..497Dolag}.

Our feedback scheme is based on the Magneticum subgrid physics model~\citep[e.g.,][]{2015ApJ...812...29Teklu}. The Magneticum AGN model is derived from \cite{2014MNRAS.442.2304Hirschmann} 
and, with a calibration similar to Magneticum suite of simulations, we set a radiative efficiency of $\epsilon_r=0.2$ to regulate the luminosity of the radiative component and we set the feedback energy per unit time to have a contribution of $\epsilon_f=0.075$ of the luminosity, as detailed in Eqs. 7-12 in \cite{2015MNRAS.448.1504Steinborn}.

\subsection{Dark matter models}

\begin{figure}
\includegraphics[width=\linewidth]{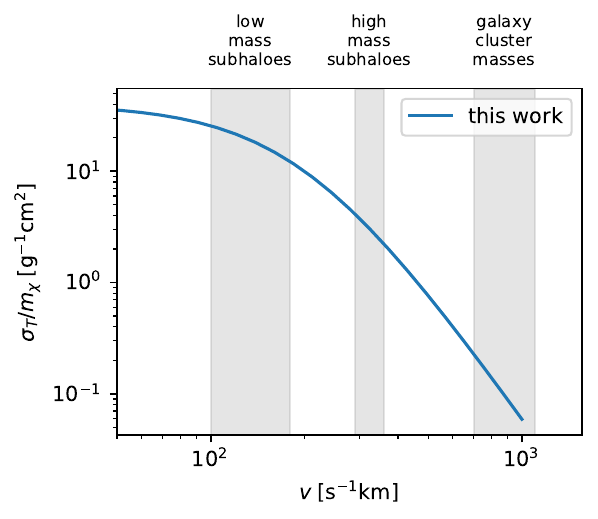}
\caption[h]{SIDM cross section as in Eq.~\eqref{eq:sigmadiv}.
Blue curve reports the cross section with the parameters chosen for this work (
$\sigma_{T0} =  40\,{\rm g}^{-1}{\rm cm}^{2}$ 
and
$v_0=200\,{\rm s}^{-1}{\rm km}$
). Vertical shaded area corresponds to the median velocity dispersion for low mass subhaloes ($M_{\rm}<2\times10^{10}\,{\rm M}_\odot$) in cluster cores, high mass subhaloes ($M_{\rm}>4\times10^{11}\,{\rm M}_\odot$) in cluster cores, and the typical velocity dispersion for haloes with $M_{\rm vir}\in[2-10]10^{14}\,{\rm M}_\odot.$ }
\label{fig:sigmadiv}
\end{figure}

As mentioned in the introduction, in addition to collisionless DM, in this work we will explore the SIDM models implemented in~\cite{Fischer2023}, which have the advantage of being capable of simulating interactions in the limit of very anisotropic cross sections and of consistently conserving the total energy of the system. 

In our SIDM models, DM interactions occur via a light-scalar mediator or a Yukawa potential. An approximation of the scattering cross section, derived using the Born approximation, yields a momentum transfer cross section $\sigma_T$ expressed as
\begin{equation}
 \frac{\sigma_T(v)}{m_\chi} = \frac{\sigma_{T0}}{m_\chi} \left(1+\left(\frac{v}{v_c}\right)^2\right)^{-2},
\label{eq:sigmadiv}
\end{equation}
Here, $m_\chi$ is the dark matter particle mass, $v$ represents the relative velocity between particles, $\sigma_{T0}$ denotes a low-velocity plateau normalisation, and $v_c$ represents the knee position before a $v^{-4}$ dependency of the cross section comes into play.

We choose the parameters $\sigma_{T0}$ and $v_c$  in order for the cross section to impact mainly low mass subhaloes  (thus $\sigma$ must be high for objects with mass $M_{\rm}<2\times10^{10}\,{\rm M}_\odot$).
Moreover we want the cross section to be relatively small for high mass subhaloes (for masses as $M_{\rm}>4\times10^{11}\,{\rm M}_\odot$, as we are not interested in modifying their properties), and it has to be negligible at cluster scales.

Given these constraints we decided to use the functional form of Eq. \eqref{eq:sigmadiv} with $\sigma_{T0}/m_\chi= 40 \,{\rm cm}^{2}{\rm g}^{-1}$ and $v_c = 200\,{\rm s}^{-1}{\rm km},$  
similar to the constraints from dwarf spheroidal galaxies proposed in ~\cite{Correa2021vdSIDM}.
In Fig.~\ref{fig:sigmadiv} present our velocity dependent cross section, together with the relevant mass scales.
Here it is easy to appreciate the decreasing importance of the cross section for increasing object masses: small subhaloes ($v\approx100\,{\rm s}^{-1}{\rm km}$) will be greatly impacted by SIDM ($\sigma/m_\chi\approx40 \,{\rm cm}^{2}{\rm g}^{-1}$), while galaxy clusters ($v\approx1000{\rm s}^{-1}{\rm km}$) will be only slightly affected by SIDM ($\sigma/m_\chi\lesssim0.1 \,{\rm cm}^{2}{\rm g}^{-1}$).

\subsection{Halo Sample}

\begin{figure*}
\includegraphics[width=\linewidth]{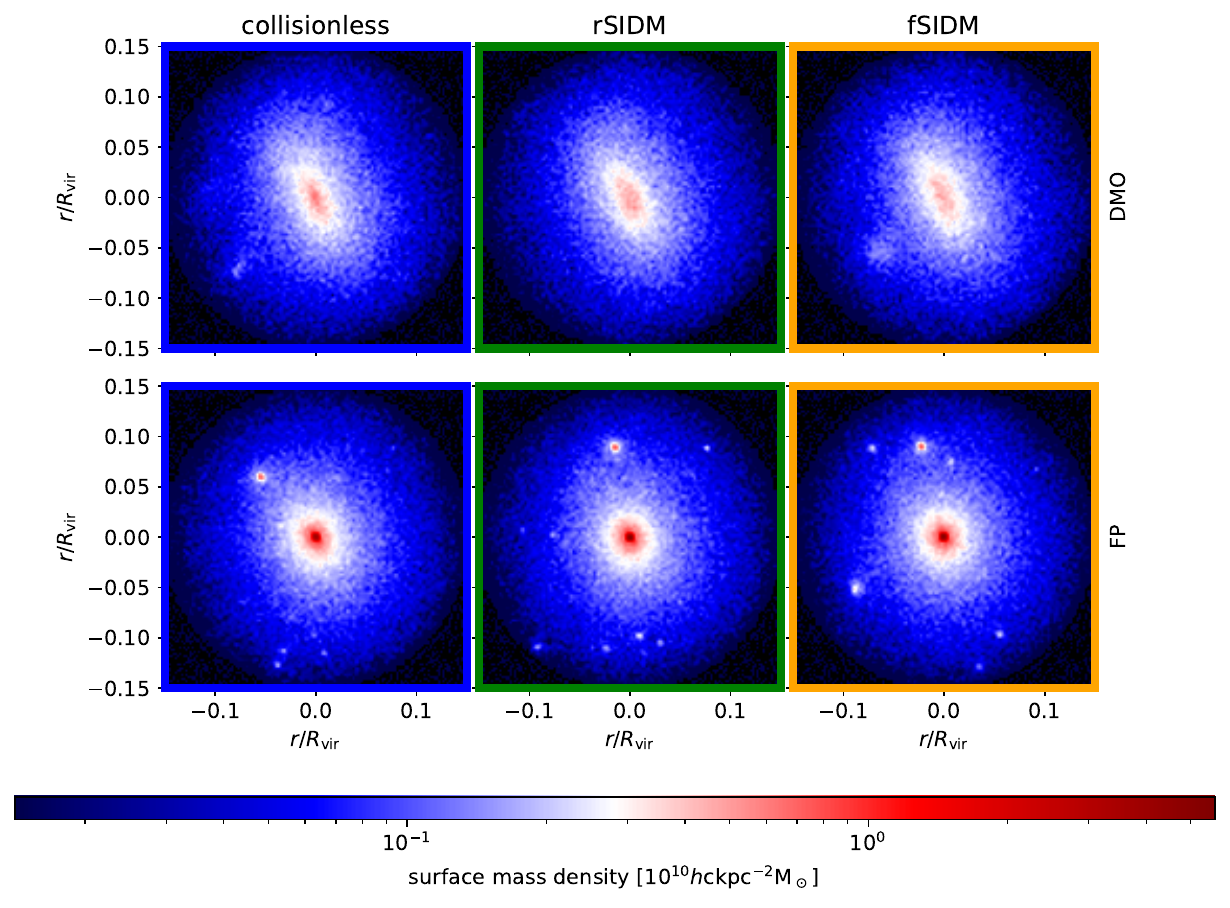}
\caption[h]{Projected density maps of central region (within $r<0.15R_{\rm vir}$) of the D16 region  re-simulations at our lowest redshift slice of $z=0.2.$
The upper row shows DMO simulations, while the bottom row refers to FP simulations. 
Columns from left to right show: collisionless DM simulations, rSIDM simulations, and fSIDM simulations, respectively.
}
\label{fig:maps}
\end{figure*}

We re-simulate the initial conditions of the regions  with DMO and FP simulations and with three dark matter types: collisionless DM, rSIDM (in our case referring to an isotropic cross section), and fSIDM,
which means that each galaxy cluster is resimulated six times.
We report the main characteristics of our galaxy cluster sample masses and concentration at the lowest redshift available ($z=0.2$) in Table \ref{tbl:sims}.
Here we estimate the dynamical state of haloes by computing the so called concentration parameter~\citep[which is known to be a good proxy for it, see e.g.,][]{2012MNRAS.427.1322Ludlow}, with objects with higher concentration being mostly early-formed and more relaxed. 
We compute the concentration parameter by assuming a Navarro--Frank--White~\citep[NFW,][]{1997ApJ...490..493Navarro} profile $\rho_{\rm NFW}$ so that
\begin{equation}
\label{eq:nfw}
\rho_{\rm NFW}\left(r\right) = \frac{\rho_0}{(r/r_s)(1+r/r_s)^2},
\end{equation}
where $\rho_0$ and $r_s$ are free parameters and the concentration is  defined as $c\equiv R_{\rm vir}/r_s.$\footnote{
The fit is performed by minimising the sum of squared log-residuals over $20$ logarithmically spaced bins in the range $[0.01,1]R_{\rm vir}.$}
We see that our haloes have concentrations in the range $c\in[2.8,5.5]$ ~\citep[we used the predictions from][]{Ragagnin2021MCcosmi}. Note that region D5 has a higher than average concentration implying that it is a relaxed system~\citep{2012MNRAS.427.1322Ludlow}.
We report the concentration for both the DMO and FP runs as they are known to differ~\citep[see e.g.][]{2010MNRAS.405.2161Duffy}, with DMO concentration being slightly lower than FP concentration.
For what concerns the concentration difference between collisionless DM and SIDM, we found  no significant changes, as expected SIDM  affect the matter distribution mainly on small scales.

Figure \ref{fig:maps} shows the projected density maps of the central region (where SIDM has most impact) of the most massive region (D16) at $z=0.2$ for the corresponding six re-simulations (three DM models with and without baryon physics).
Moving from DMO to FP simulations we see much more peaked substructures and that in general SIDM models tend to have substructures that are displaced differently compared to the CDM model, due to the expected different trajectories during mergers~\citep[as shown in][]{Sabarish}.

\section{Galaxy cluster matter density profiles}
\label{sec:profi}

\begin{figure}
\includegraphics{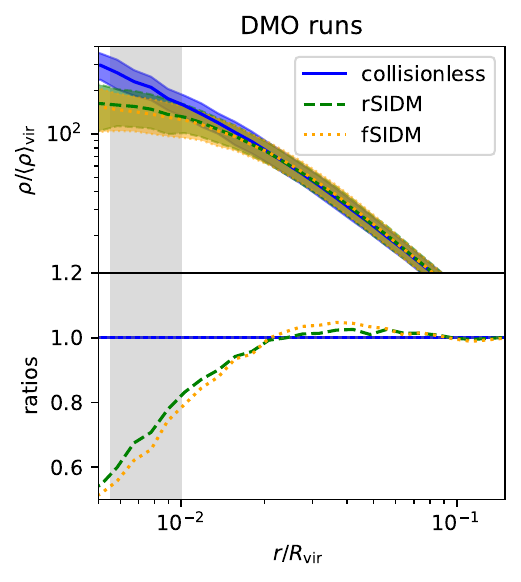}
\caption[h]{Comparison between stacked central density profile in the  DMO runs. We report collisionless DM density profile as blue solid line, rSIDM as green dashed line, and fSIDM as orange dotted line (note that rSIDM and fSIDM lines almost overlap visually). Shaded areas correspond to one standard deviation of the six regions. Upper panel shows the central density, normalised by the average density at $R_{\rm vir}$ for our six regions,  while the bottom panel shows the ratio between the profiles of the  SIDM runs and the collisionless DM runs. Gray shaded area corres one standard deviation around the fractional fiducial DM softening $2.8\times\epsilon/R_{\rm vir}.$}
\label{fig:projeDMO}
\end{figure}

\begin{figure*}
\includegraphics[width=\linewidth]{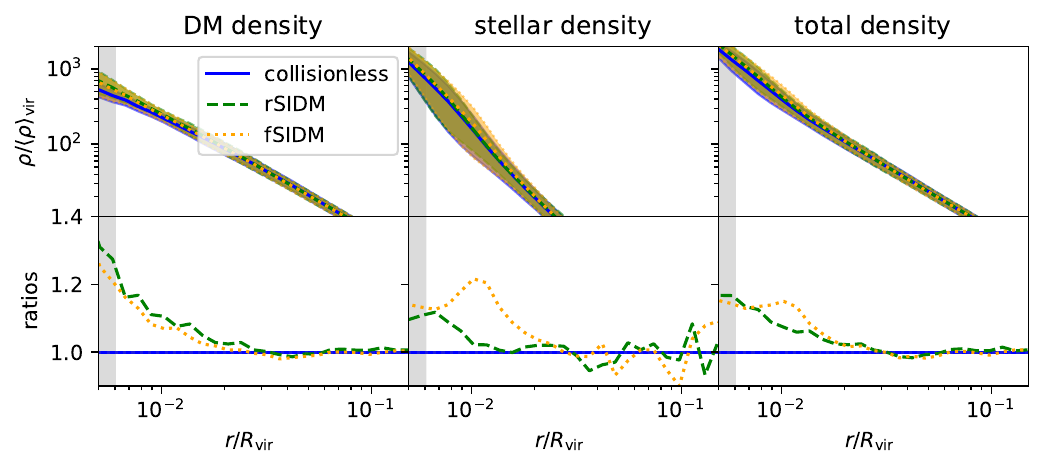}
\caption[h]{Comparison between stacked central density profiles in the FP runs. DM models are represented as in Fig. \ref{fig:projeDMO}. Columns refer to the results for DM, stellar, and total matter components respectively. Upper panels show DM density profiles, while lower panels show the ratio with respect to the collisionless DM profiles. Gray shaded area represents one standard deviation around the fractional fiducial stellar softening $2.8\times\epsilon_\star/R_{\rm vir}.$} 
\label{fig:projeFP}
\end{figure*}

In this section, we will concentrate on the distribution of matter and subhaloes within the cluster cores, specifically within a cluster-centric distance $r<0.15R_{\rm vir}.$
The reason behind this choice is twofold: on these small scales the effect of SIDM is expected to be largest, as it is a region sensitive to the nature of DM models~\citep[see, e.g.,][]{2007MNRAS.376..180Natarajan,2021PhRvD.104j3031Yang}; moreover, this is the typical field of view of galaxy-galaxy strong lensing analyses from Hubble Frontier Field data~\citep[see, e.g.,][]{2019A&A...631A.130Bergamini,2022A&A...659A..24Granata,Granata2023}.

Figure \ref{fig:projeDMO} presents the stacked density profiles within the central regions of our DMO simulations, encompassing redshift slices between $0.2<z<0.6.$ Given the broad virial mass range across our six regions, we opted to stack all haloes together by rescaling their profiles using the virial radius and virial mass.
As expected SIDM simulations yield cored profiles~\citep[see e.g.,][]{2020PhRvD.102d3009Kamada}, in particular our SIDM DMO simulations form a core at radii $r<2\times10^{-2}R_{\rm vir}.$ 
The ratio between collisionless and self-interacting DM (see bottom panel of Fig.~\ref{fig:projeDMO}) becomes significantlt different from unity at cluster centric distances $\lesssim2\times10^{-2}R_{\rm vir}$ (approximately corresponding to a cluster-centric distance of $\approx 100{\rm kpc}$).
It is worth noting that the   profiles are different at distances larger than the fiducial DM softening ($2.8\epsilon_{\rm DM}$,  indicated by the grey vertical band).
Consequently we can conclude that in DMO SIDM simulations, the impact extends across the entire cluster 
central regions
with both  SIDM variants (rSIDM and fSIDM) exhibiting $\approx20\%-40\%$ lower density compared to collisionless DM.

We present  results for the FP simulations in Fig. \ref{fig:projeFP} (top row), where the stacked density profiles of DM, stellar, and total density for our six Dianoga regions are shown. Notably, in SIDM FP simulations, both rare and frequent scenarios tend to exhibit DM central densities that are more pronounced than those observed in the corresponding FP collisionless DM simulations.
Notice that \cite{Robertson2019SIDM+baryons} reported an opposing result, demonstrating that their hydrodynamic SIDM simulations produced cluster cores that, while less cored than the DMO run, still retained a degree of core structure compared to the collisionless run.
The reason for this difference remains unclear,  however we can speculate that since the baryonic central potential is known to impact the central density evolution~\citep[see e.g.,][]{Elbert2018SIDMAndBaryons}, the different results may be attributed to different central baryon density in our study compared to that in~\cite{Robertson2019SIDM+baryons}.

To verify this statement we will conduct the following two analyses: (I) in order to ensure to have our FP collisionless DM profile under control, in Sect. \ref{sec:adiab} we will first verify that the increased DM cuspiness in collisionless  DM runs is due to adiabatic contraction; then (II) in Sect. \ref{sec:cc} we will compare our SIDM central densities with analytical models of  gravothermal solutions of SIDM haloes in the literature to understand if our increased central density agrees with gravothermal evolution models.

\subsection{Adiabatic Contraction}
\label{sec:adiab}

\begin{figure}
\includegraphics[width=1\linewidth]{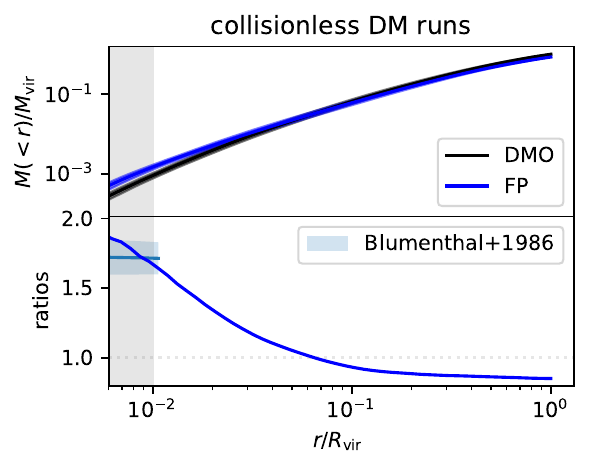}
\caption[h]{Comparison between DM mass profiles of collisionless DM runs. Top panel shows the stacked DM mass profile of DMO runs (black) and FP runs (blue), while bottom panel shows their ratio and the central value predicted by the adiabatic contraction model presented in ~\cite{Blumenthal1986AC} as light-blue shaded blue area. The vertical gray shaded area corresponds to one-standard deviation of the DM gravitational softening.}
\label{fig:contra}
\end{figure}

FP simulations (compared to DMO ones) produce a heightened concentration of DM close the cluster centre~\citep[as in][]{2020MNRAS.491.1295Despali} due to adiabatic contraction~\citep[as found above and for example in][]{Gnedin2004AdiabaticContraction}, 
 where the baryon density in halo cores is high enough to generate a back reaction on the dark matter component and increase its central density significantly
 compared to DMO  runs, see for instance Fig. 5 in \cite{2010MNRAS.405.2161Duffy}.
 
 In this subsection we will verify that we have under control the results of our collisionless DM simulations,  we will use only collisionless DM runs (both FP and DMO), and estimate the  contribution to the central baryon density from adiabatic contraction following the theoretical model proposed by ~\cite{Blumenthal1986AC}.
They derive that the halo radius $r$ times mass within it is a constant, namely
\begin{equation}
r M_{\rm dm,i}(<r) = r\left[M_{\rm dm,f}(<r) + M_{\rm b}\right],
\end{equation}
where $M_{\rm dm,i}$ is the initial dark matter mass, $M_{\rm dm,f}$ is the final dark matter mass and $M_{\rm b,f}$ is the final baryon mass.
In our work, we estimate $M_{\rm dm,i}$ as the dark matter mass from DMO runs, $M_{\rm dm,f}$  as the dark matter mass in the FP runs, and $M_{\rm b,f}$ as the baryon mass in the same FP runs. 
We cross-matched the DMO and FP masses of each halo in all available snapshot and computed the adiabatic contraction factor using the software \texttt{contra}~\citep{Gnedin2004AdiabaticContraction}.

We report our finding in Fig.~\ref{fig:contra} (top panel), where we show the stacked DM mass profile of our haloes from collisionless DM simulations from both the DMO and FP runs.
The bottom panel of  Fig.~\ref{fig:contra} reports their ratio together with the median prediction from the model by~\cite{Blumenthal1986AC} for each of the cross-matched haloes (the size of the shaded area represents the error on the mean).
We can see that the model by~\cite{Blumenthal1986AC}  correctly predicts the increased central density of the FP simulations, and we therefore confirmed that the increased DM density found in our collisionless FP (compared to collisionless DMO) simulations is due to adiabatic contraction produced by  baryons.

\subsection{Gravothermal Solution}
\label{sec:cc}

\begin{figure}
\includegraphics[width=1\linewidth]{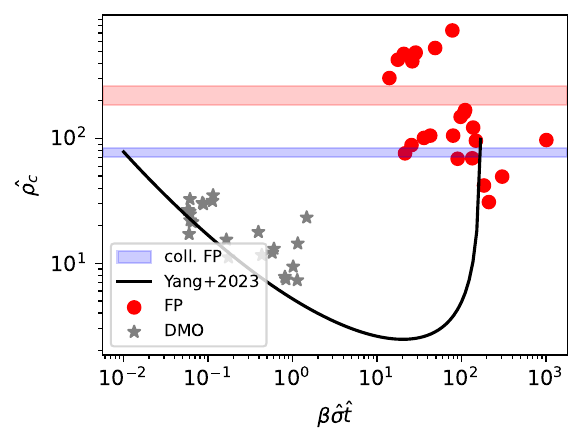}
\caption[h]{Rescaled DM central density $\hat \rho_c$ as a function of rescaled time $\beta\hat \sigma\hat t,$ for our FP SIDM (red circles) and DMO SIDM (gray stars) runs. We also overplot the model from \citep[][black solid line]{Yang2023CC}. Red shaded area shows the mean value of the FP SIDM points with its one standard deviation, while blue shaded area shows the same for collisionless FP runs.}
\label{fig:collapse}
\end{figure}

\begin{figure}
\includegraphics[width=1\linewidth]{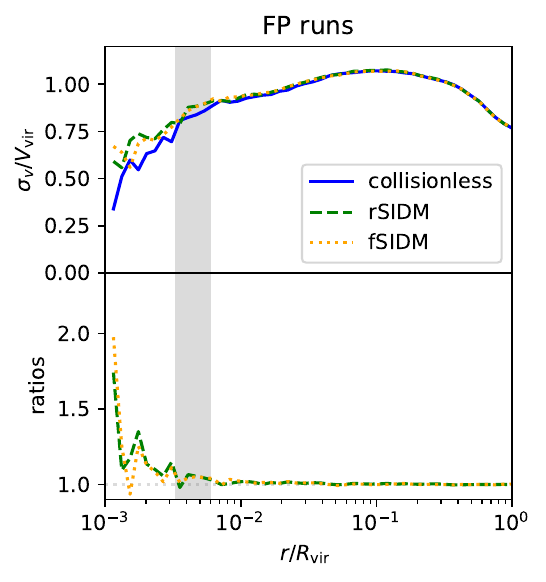}
\caption[h]{Comparison between stacked velocity dispersion profiles. Top panel presents the  stacked velocity dispersion profiles for our FP simulations galaxy clusters in units of virial velocity $\sqrt{GM_{\rm vir}/R_{\rm vir}}.$
Bottom panel: ratios with respect to the DMO counter parts. Lines are coloured as in Fig.~\ref{fig:projeFP}.}
\label{fig:profivel}
\end{figure}

We will compare the central density of our haloes with the gravothermal solution for SIDM haloes proposed in \cite{Yang2023CC}. Their model is based on the assumption that evolution of central density and core radius of SIDM NFW haloes can be described by a single parameter $\beta$ once variables are rescaled as follows: $t_0=1/\sqrt{4\pi G\rho_s},$ radii by factor $r_0=r_s,$ densities by a factor $\rho_0=\rho_s,$ and cross sections by a factor $\left(\sigma/m_\chi\right)_0 = 1/(\rho_s r_s).$ We denote rescaled quantities with a hat.

We then estimate the core size of our DM haloes of SIDM simulations using the DM profile proposed in their Eq. A5, that is a simil-NFW profile with a possible core radius $r_{\rm c}$ and a truncation $r_{\rm out},$ namely
\begin{equation}
\label{eq:core}
\hat\rho\left(\hat r\right) = 
\frac{\hat\rho_{\rm c}}
{1+(\hat r/r_{\rm c})^s(1+\hat r/r_{\rm out})^{3-s}},
\end{equation}
 where $\hat\rho_c$ is the re-scaled central density, and we use $s=2.19$ as proposed by \cite{2021PhRvD.104j3031Yang}.

Note that the time rescaling factor $t_0$ proposed by  \cite{Yang2023CC} does not take into account that FP SIDM  simulations have a much shorter evolution time compared to DMO ones~\cite[as shown in][]{Zhong2023}.
Therefore, for FP simulations we follow the approach of ~\cite{Zhong2023} and we rescale the time variable by combining Eqs. 15 and 16 in ~\cite{Zhong2023}, namely by a factor of $\left(\rho_{\rm eff}\sqrt{\left|\Phi(0)_{\rm FP}\right|}\right)/\left(
    \rho_s\sqrt{\left|\Phi(0)_{\rm DMO}\right|}\right),$
where $\Phi(0)_{\rm DMO}$ and $\Phi(0)_{\rm FP}$ are the central gravitational potential, for the DMO and FP counterpart of a halo, and $\rho_{\rm eff}$ is an effective density that captures the contraction effect due to the baryonic potential.
To this end, for each halo we estimate $\rho_{\rm eff}$  as the central density of baryons and we
estimate it using their Eq. 17, namely $\rho_{\rm eff} \approx \rho_s + \alpha M_{\rm b}/(2\pi r_{\rm s} r_{h}),$ where $M_{\rm b}$ is the baryon mass of the halo and  they set $\alpha=0.4.$ Following their procedure, we compute $r_h$ by fitting an Einasto~\citep{hernquist90} profile to the baryonic component (see Appendix~\ref{ap:evo} for the fit the details). 

We then estimate the cross section acting on the galaxy cluster core by computing the effective cross section $\sigma_{\rm eff}$ similarly to~\cite{Yang_2022D}, where the average  cross-section is weighted with fifth power of the velocity and assuming that the velocities are well described by a Maxwell–Boltzmann distribution (with a characteristic velocity $V_{\rm max}/\sqrt{3},$ see their  Eq. 4.2):
\begin{equation} \label{eq:sigma_eff}
    \sigma_\mathrm{eff} = \frac{3}{2} 
    \frac{\langle v^5 \sigma_v(v) \rangle}{\langle v^5 \rangle} \,,
\end{equation}
where $\sigma_v$ is the viscosity cross section given by
\begin{equation} \label{eq:sigma_v}
    \sigma_v = 4\pi \int_0^1 \frac{\mathrm{d}\sigma}{\mathrm{d}\Omega} \sin^2 \theta \, \mathrm{d}\cos\theta \,.
\end{equation}

We compute the rescaled time variable $\beta\hat \sigma\hat t$ as proposed in ~\cite{Yang2023CC}, where $\beta$ is the free parameter of their model that should vary around one, therefore we keep it as $\beta=1$ for simplicity.
In Fig.~\ref{fig:collapse}  we  compare our DM central densities $\hat \rho_c$ for both the DMO (gray stars) and FP (red circles) SIDM runs with the prediction of $\hat \rho_c\left(\beta \hat \sigma \hat t\right)$    from ~\cite{Yang2023CC}.
We can see that DMO $\hat \rho_c$  lies in the core-formation phase of  ~\cite{Yang2023CC} central density evolution, namely in the left part of the plot.

For what concerns the FP simulation, we present mean values (with their one standard deviation) for $\hat \rho_c$ in the case of SIDM (red shaded area) and collisionless FP (blue shaded area) simulations.
The mean value of $\hat \rho_c$ for collisionless FP simulations exceeds the corresponding value in the case of DMO, consistently with expectations from adiabatic contraction. 
On the other hand, the SIDM FP values of $\hat \rho_c$ are higher than the central densities found in collisionless DM FP simulations.
This increase is due to the baryonic potential even if the haloes are not in a core-collapse phase.
In support to this hypothesis, we illustrate the velocity dispersion profiles of our clusters in Fig.~\ref{fig:profivel}, which reveals that the central region of SIDM clusters exhibits an increasing profile, therefore even if the profile is less steep than their collisionless FP counterpart, the FP SIDM clusters are still in their core formation phase (see Appendix~\ref{ap:evo} for more details on the density and velocity dispersion time evolution).

\section{Subhalo population}
\label{sec:sh}

\begin{figure*}
\includegraphics[width=\linewidth]{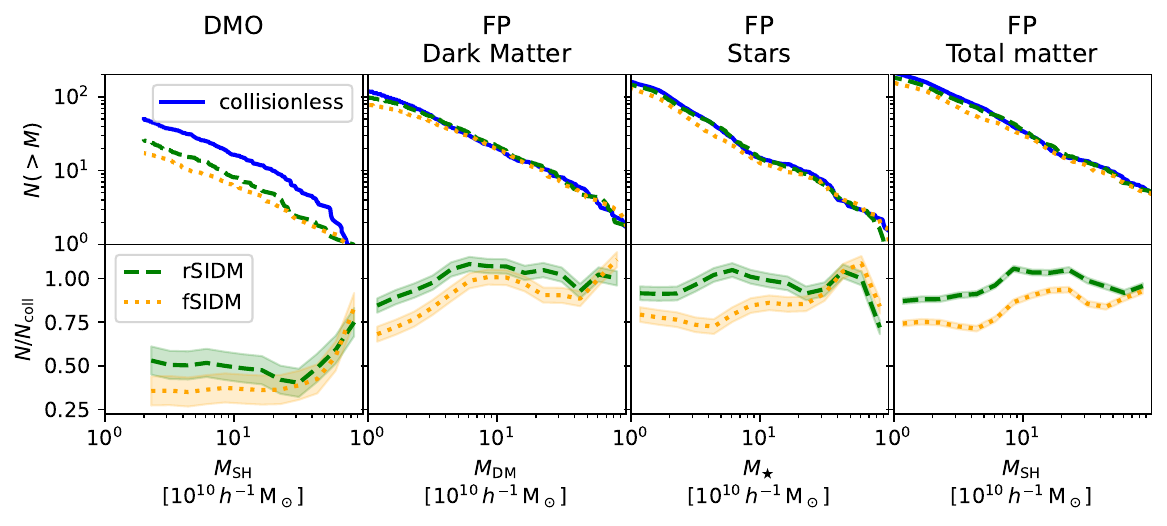}
\caption[h]{Cumulative SHMFs in units of the virial mass for subhaloes   with a cluster centric distance $<0.15R_{\rm vir}.$ The left most panel focus on DMO run with collisionless DM and rare and frequent SIDM; the other panels refer on FP simulations and show the results for different mass components: from left to right, total matter, stellar matter and DM. Colours and line styles are as in Fig. \ref{fig:projeDMO}.
The panels in the bottom row show the ratio between the SHMFs in the SIDM models and the one for the collisionless DM model.
Shaded area corresponds to one standard deviation of the mean value.
}
\label{fig:shmf}
\end{figure*}

In this Section, we analyse the subhalo population in the central regions of our simulated clusters. In particular, we focus first on comparing SIDM subhalo abundances, then we explore whether these values agree with theoretical models on tidal stripping, and finally, we study subhalo compactness as this has potential value for future lensing analyses.

\subsection{Subhalo abundance}
\label{sec:shmf}

Our subhalo abundance analyses are specifically focused on subhaloes within a cluster-centric distance of $<0.15R_{\rm vir}$ as the central part of the cluster is the one most affected by SIDM properties.
Moreover, the central region of clusters is also relevant for strong lensing investigations, as highlighted in~\cite{Meneghetti2023PersistentExcess}.
However, more precise comparisons are needed in the future, for instance, considering projection effects and precise computation of the reduced shear to compare simulated and observed data correctly.

In the top panel of Fig. \ref{fig:shmf} (top panel), we present the cumulative Subhalo Mass Function (SHMF) for our sample at low redshift in the  DMO  simulations. Notably, there is a pronounced suppression of low-mass subhaloes in these simulations.
Satellite suppression is expected in SIDM simulations because of the enhanced self-interactions of satellites with the host~\citep{Nadler2020SIDM}.
In the remaining top panels of Fig. \ref{fig:shmf}, we depict the FP  SHMF and dissect it based on the different matter components, including stars and DM. The relative values of the cumulative mass function are presented in the bottom panels of Fig. \ref{fig:shmf}, revealing that SIDM has a smaller impact on the subhalo population in FP simulations compared to DMO simulations.
The suppression of satellites in DMO simulations can be substantial, reaching approximately $\approx50\%$ for rSIDM runs and going as high as $\approx65\%$ for fSIDM simulations. In FP simulations, the suppression of satellites is more moderate, though still significant, with approximately $\approx 10\%$ for rSIDM and $\approx 35\%$ for fSIDM in terms of total masses.
An increased suppression of satellites in fSIDM models compared to rSIDM is expected; see, for instance, \cite{2022MNRAS.516.1923Fischer} and   \cite{Fischer2023}. 
In general, the smaller satellite suppression produced by SIDM in FP simulations is due to the presence of a stellar component that makes subhaloes more resistant to disruption.

\subsection{Tidal stripping}

\begin{figure*}
\includegraphics[width=1.\linewidth]{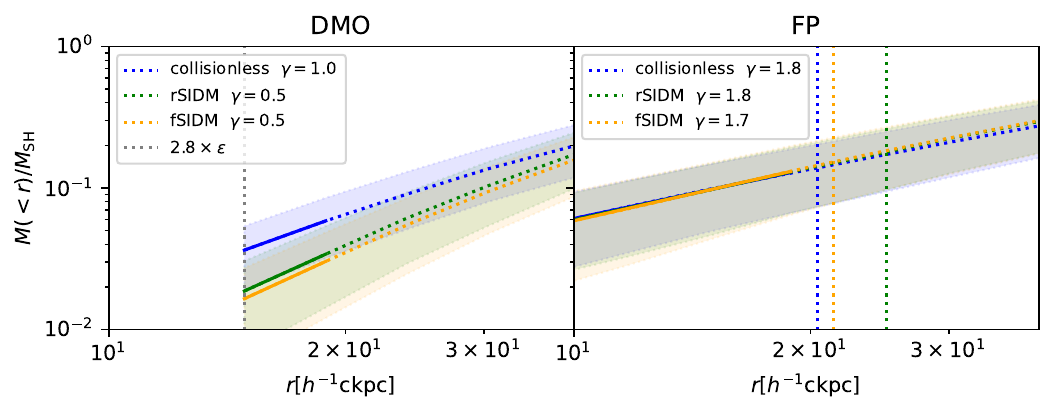}
\caption[h]{Stacked cumulative matter profiles of subhaloes in the outskirt of galaxy clusters (outside $r>0.5R_{\rm vir}$), for DMO (left panel) and FP (right panel) simulations for collisionless (blue dotted lines), rSIDM (green dashed line), and fSIDM (orange dotted line) models. 
Solid lines show the profile fitted in the internal region ($r<20\,{\rm ckpc}$). Vertical dashed lines correspond to the mean half-mass radius.}
\label{fig:gamma}
\end{figure*}

We now investigate the subhalo suppression that we found in SIDM simulations in the light of other theoretical studies.
In particular, we want to test if this suppression of substructures is consistent with theoretical works as
~\cite{Penarrubia+2010}, which shows that the suppression details are strongly dependent on the inner log-slope $\gamma$ of their matter profiles.

In the left panel of  Fig.~\ref{fig:gamma}, we present the stacked cumulative density profiles of subhaloes in the outskirt of galaxy clusters (outside $r>0.5R_{\rm vir}$) for our three DM models in DMO simulations, that we assume to be in an infalling phase. We fit the internal log-slope of the total matter profile and assume it equals to $\left(3-\gamma\right)$ between the softening and $20\,{\rm ckpc}.$
We opted for this radial range because it is large enough to capture the internal log-slope and small enough not to encounter the flattening in the tail of the profile.
To show that our radial range is small, we over-plot the mean half mass radius as vertical lines, where we can notice that they are larger than  $20\,{\rm ckpc}$).
We can see that the DMO collisionless run has  $\gamma\approx1,$ which corresponds to an NFW-like core, as expected from theory.
The DMO SIDM runs have a significantly lower profile, with fSIDM having a flatter profile than rSIDM, in agreement with the SHMF that we presented in Fig.~\ref{fig:shmf}, where fSIDM has a larger suppression than rSIDM.

In the right panel of  Fig.~\ref{fig:gamma}, we present the same results but for FP simulations.
Our  $\gamma$ values correspond to a cusp (as expected, baryons dominate the central part) and both SIDM and collisionless runs have very close values that are in agreement with the much milder SIDM suppression of satellites that we found in the SHMF shown in Fig.~\ref{fig:shmf}. 

It is possible that the significant subhalo suppression of fSIDM with respect to rSIDM in FP simulations is due to a different effect than tidal stripping; 
in fact, the two models have similar density profiles (see Fig.~\ref{fig:gamma}).
Therefore, one can speculate that the different SMHFs could be attributed to the differences in the host potential~\citep{Penarrubia+2010}; the fact that the tests performed in ~\cite{Penarrubia+2010} assume a static host halo; or the effect of dark matter self-interactions between the infalling subhalo and the host.

\subsection{Subhaloes compactness}
\label{sec:compact}

\begin{figure*}
\includegraphics[width=\linewidth]{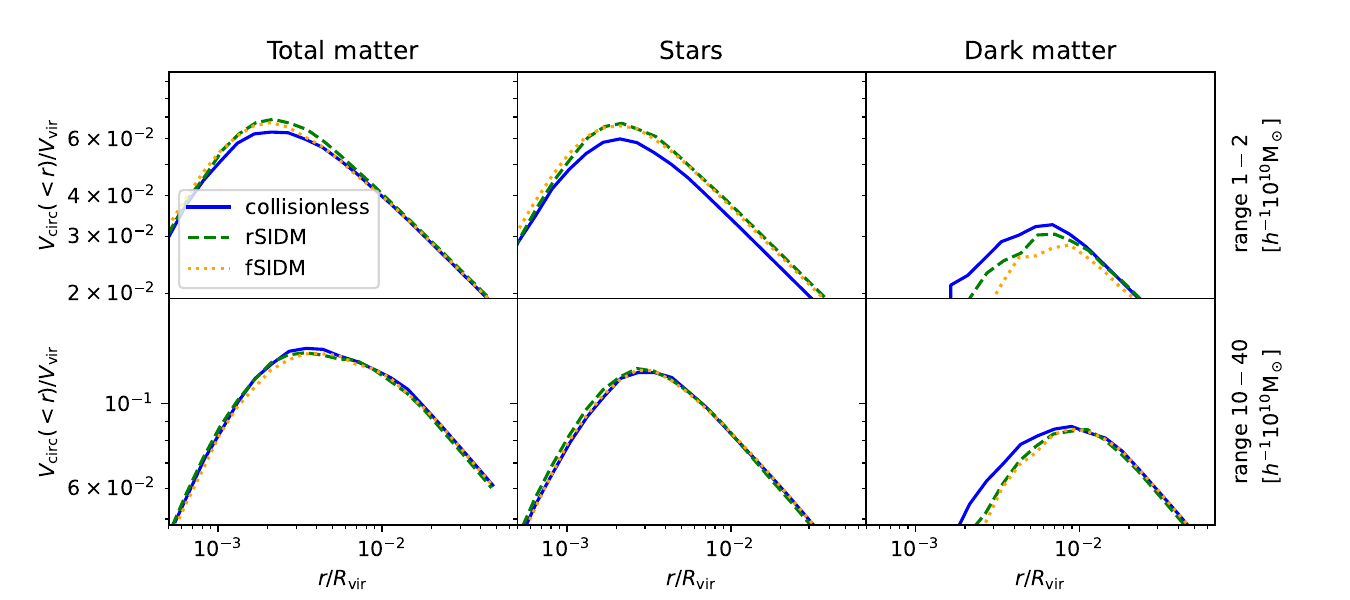}
\caption[h]{ Circular velocity profiles for FP simulations with a cluster centric distance $<0.15R_{\rm vir}$. Each column shows a different matter component: total matter on the left, stellar component in the middle and the DM component on the right. Top panel: low mass regime $[1,2]h^{-1}10^{10}{\rm M}_\odot$, bottom panel: high mass regime $[10,40]h^{-1}10^{10}{\rm M}_\odot$. Line styles are as in Fig.~\ref{fig:projeDMO}.}
\label{fig:Vcircprof}
\end{figure*}

 \begin{figure*}
\includegraphics[width=\linewidth]{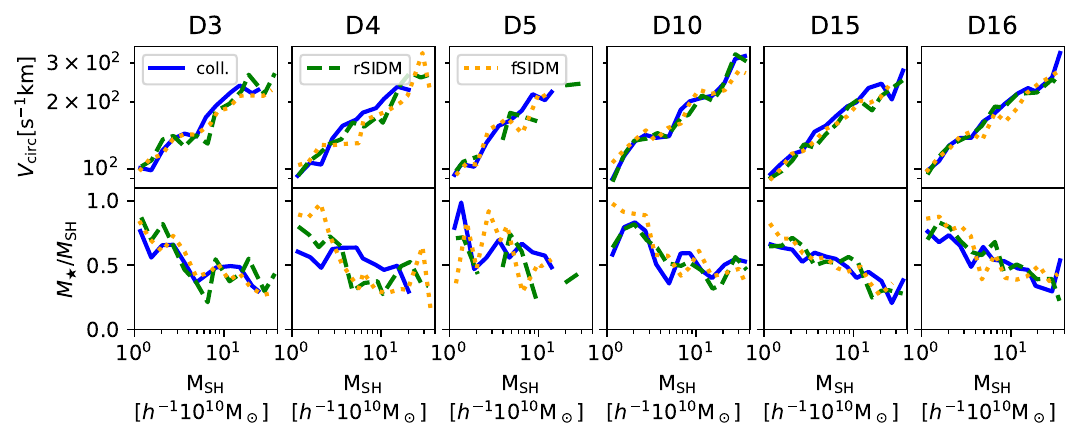}
\caption[h]{Subhalo compactness and stellar fraction for subhaloes for cluster centric distances $<0.15R_{\rm vir}$. The upper panels show the average circular velocity per subhalo mass bin, while the bottom panels refer to the stellar fraction. Each column reports the results for a given re-simulated region (see Table~\ref{tbl:sims}). Colour codes and line styles are as in Fig. \ref{fig:projeDMO}.}
\label{fig:bergaplot}
\end{figure*}

We will now investigate how SIDM impacts the compactness of subhaloes in galaxy cluster cores, as it is the region that is most affected by SIDM.
The work of ~\cite{Penarrubia+2010} shows that cuspy subhaloes (as the ones we produce in our FP runs) will lose much more mass and decrease circular velocity only slightly (see the upper panel of their Fig. 6).
As central substructures are supposed to experience earlier infall, it is expected that they would have undergone substantial dark matter loss while maintaining a circular velocity closely resembling that at the time of infall. The net consequence of this mechanism is an increased circular velocity at fixed subhalo mass.

To verify this, we present the circular velocity profiles in Fig.~\ref{fig:Vcircprof}.
We observe an increase in the circular velocity of SIDM $V_{\rm circ}$ (up to approximately $\approx20\%-30\%$) with respect to the collisionless DM $V_{\rm max}$.
In order to understand the mechanism that increased the compactness of these low-mass subhaloes, we dissect them and analyse the circular velocity profiles of their stellar (central column) and DM component (right column). 
Here, it is clear that the circular velocity stellar component of the SIDM runs is higher than that of collisionless DM runs.
To prove this point better, in Fig. \ref{fig:bergaplot}, we show the subhalo maximum of circular velocity (upper panels) and stellar fraction (lower panels) versus $M_{\rm SH}$ mean relations for each of our six re-simulated regions 
  
  Therefore, we confirm that the increase in the circular velocity of subhaloes in SIDM simulations is associated to an increase in the stellar fraction.
  In particular, tidal stripping lowers the total mass while the circular velocity decreases only slightly, therefore bringing the data points closer to the observational scaling relation from ~\cite{2019A&A...631A.130Bergamini}.
  
It is interesting to note in  Fig.~\ref{fig:Vcircprof} that SIDM can produce subhaloes with higher compactness in the low-mass regime of subhaloes. This result reduces the tension between observed and simulated low-mass subhalo compactness~\cite{2022A&A...665A..16Ragagnin} without changing baryon physics.

\begin{figure}
\includegraphics[width=1.\linewidth]{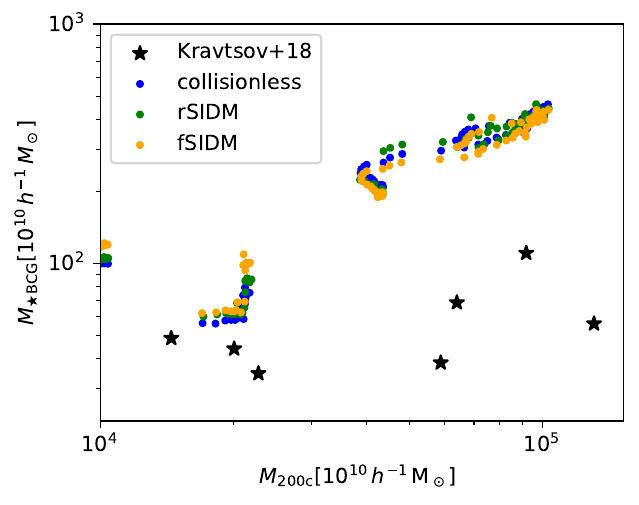}
\caption[h]{Stellar mass of the BCG as a function of the halo mass $M_{\rm 200c}$.  Colour data points are as in Fig. \ref{fig:projeDMO}. Black stars represent points from \cite{2018AstL...44....8Kravtsov}. }
\label{fig:Kratsovplot}
\end{figure}

We now tackle the problem of the BCG stellar masses being overly massive compared to observations.
To this end, in Fig. \ref{fig:Kratsovplot} we show the BCG stellar mass $vs.$ the total mass of the halo, where we see that the mass of the BCG is only slightly affected by the inclusion of SIDM models. This is mainly related to the BCG being supposedly accreted by ex-situ material~\citep[see, e.g.,][]{2023MNRAS.521..800MontenegroTaborda}.
If subhaloes formed their stars in the field, then the subhalo stellar mass would be poorly influenced by SIDM.

\subsection{Subhaloes with low circular velocity}
\label{sec:fi}

 \begin{figure}
\includegraphics[width=\linewidth]{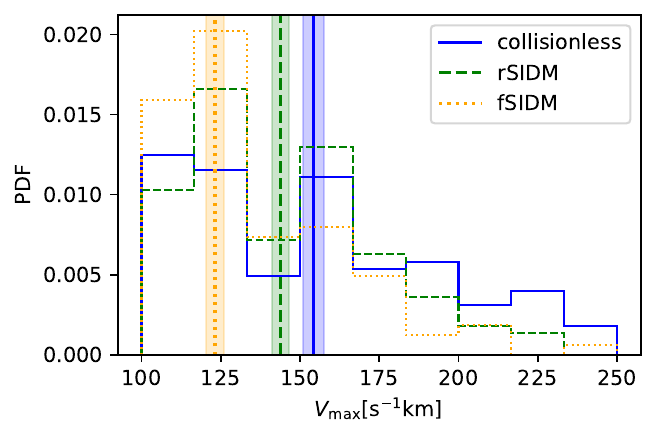}
\caption[h]{Peak of circular velocity for all galaxies with mass cut of $1<M_{\rm SH}<2\,[h^{-1}\,10^{11}\,{\rm M}_\odot]$ for subhaloes with cluster centric distance in the range $[0.15,1]R_{\rm vir}.$ 
 Colour lines and styles are as in Fig. \ref{fig:projeDMO}. Shaded vertical bands correspond to the median and one standard deviation on the mean.  }
\label{fig:vcirc_fuori_all}
\end{figure}

 \begin{figure}
\includegraphics[width=\linewidth]{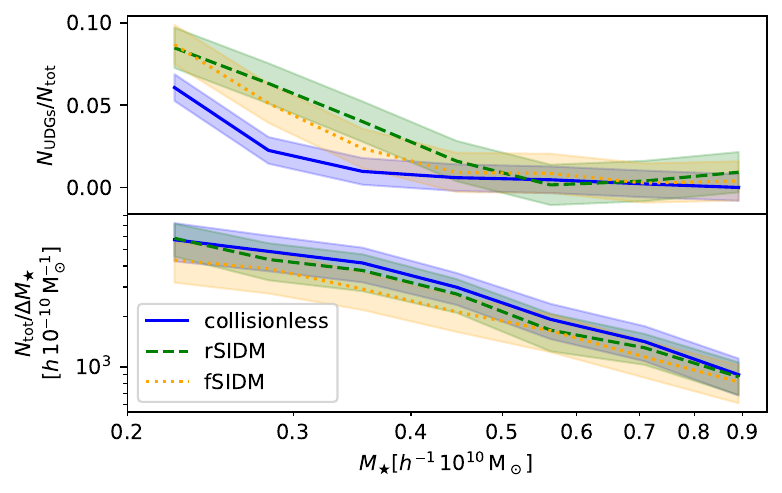}
\caption[h]{Number of objects per mass stellar mass bins (upper panel) and the fraction above $\Sigma_{\rm UDG}$ defined in \cite{Sales+2020UDGplot} for a given stellar mass bin.
Here we consider only objects outside the main halo $R_{\rm vir}$ and inside the main halo FoF group.
Colours are organised as in Fig. \ref{fig:projeDMO}, the shaded area corresponds to the error of the mean in each bin. }
\label{fig:udgh} 
\end{figure}

In this section, we examine how SIDM models impact the low-concentration tail of subhaloes. To investigate this, we analyse the distribution of $V_{\rm max}$ for subhaloes within the mass range of $1-2\,10^{10}h^{-1}\,{\rm M}_\odot$ and located between $0.15R_{\rm vir}$ and $R_{\rm vir}$. We exclude subhaloes in the central regions of clusters, as we are aware from Sect.~\ref{sec:compact} that stripping in this region may result in higher compactness.
In Fig.~\ref{fig:vcirc_fuori_all}, we present the PDF of $V_{\rm max}$ for these subhaloes, revealing that SIDM simulatons exhibit significantly lower median values of $V_{\rm max}$. This result is compatible with the increase in circular velocity found in DMO simulations~\citep[see Fig. 14 in][]{Fischer2023}, and in this work, we confirm that this effect is still present in hydrodynamic simulations.

We now study if SIDM models are capable of producing an increased fraction of UDGs compared to collisionless DM.
We define these objects as in \cite{Sales+2020UDGplot}, namely as systems with $M_\star/R_{\rm eff}<2.6\times10^{7}\,10^{10}\,h\,{\rm kpc}^{-2}\,{\rm M}_\odot,$ where $M_\star$ is the galaxy stellar mass and $R_{\rm eff}$ is the projected half light radius respectively ~\citep[defined as in][namely as $4/3$ times the 3D half mass radius]{Sales+2020UDGplot}.
 We also use the same resolution cut used in ~\cite{Sales+2020UDGplot}, which imposes a lower limit of $10$ DM particles per subhalo.
In the upper panel of  Fig.~\ref{fig:udgh}, we show the fraction of UDGs per mass bin.
We notice that SIDM models produce systematically more diffuse objects, as the fraction of SIDM UDGs at $M_\star\approx0.3\times10^{10}\,h^{-1}\,{\rm M}_\odot$ is more than twice as the amount in collisionless DM, in agreement with the fact that SIDM subhaloes in cluster outskirts tend to have lower circular velocities~\citep[see e.g., Fig. 6 in][]{Penarrubia+2010}.

To conclude the section, we verify if the increase of UDGs can be due to an overall increased number of substructures.  Therefore, in the bottom panel of Fig.~\ref{fig:udgh}, we show the total number of galaxies per stellar mass bin, where we can see that SIDM models have systematically significantly fewer subhaloes compared to collisionless DM (stellar masses span $8$ log-spaced bins in the range [0.2,1] $h^{-1}\,10^{10}{\rm M}_\odot$).

\section{Conclusions}
\label{sec:conclu}
In this study, we conducted a comparative analysis of the impact of rare and frequent velocity-dependent  SIDM models within the cores of massive galaxy clusters (with masses around $\approx 10^{14}-10^{15}\,{\rm M}_\odot$). Additionally, we examined the influence of incorporating baryons in these simulations.
We obtained the following main results:


\begin{itemize}
\item DMO simulations of rare and frequent SIDM models form cored galaxy cluster profiles in agreement with other theoretical studies. In contrast, SIDM FP simulations show a significantly ($\approx20\%$)  higher central matter density compared to collisionless DM.
Therefore we warn that it is not straightforward to observationally constrain SIDM cross-section by just estimating galaxy cluster cores.
\item While SIDM DMO simulation strongly suppresses subhaloes in cluster cores compared to collisionless DM, this suppression is dampened in FP simulations; however, it is still significant, corresponding to a factor $\approx20\%$ for rSIDM and $\approx35\%$ for fSIDM. 
\item We found that SIDM produces cluster cores with substructures in the low-mass regime ($<10^{11}\,h^{-1}\,{\rm M}_\odot$) that are more compact and that this increase is due to an enhanced stripping of their DM component.
\item SIDM simulations produce a significantly higher fraction of low surface brightness galaxies in cluster outskirts compared to collisionless DM.
\end{itemize}

In the future, it would be interesting to run and study cluster simulations with a resolution that is high enough to study subhalo core-collapse. 
Recent studies show that to properly resolve the core-collapse up to redshift $z=0$ is challenging, as they require a large number of particles (compared to typical cosmological simulations) and small time steps to reproduce theoretical gravothermal evolution models~\citep{2021PhRvD.104j3031Yang,Zhong2023,2024arXiv240212452P,Mace2024ConvergenceCoreCollapse};
 to properly resolve the mean free path~\citep[see discussion in][]{2024arXiv240300739F};
 and they also need a relatively high accuracy of gravitational interactions to deal with energy conservation issues of current numerical schemes~\cite{2024arXiv240300739F}.

   
\begin{acknowledgements}
We acknowledge support from the grant PRIN-MIUR 2017 WSCC32.
AR acknowledges support by  MIUR-DAAD contract number 34843  ``The Universe in a Box''. We  carried our simulations using the the INAF-Pleiadi project SIDM vs CDM allocated in the Trieste IT framework~\citep{2020ASPC..527..307Taffoni,2020ASPC..527..303Bertocco}. 
We are especially grateful for the support by M. Petkova through the Computational centre for Particle and Astrophysics (C$^2$PAP).  LM acknowledges the support from the grants PRIN-MIUR 2022 20227RNLY3  and ASI n.2018-23-HH.0. 
KD acknowledges support by the COMPLEX project from the European Research Council (ERC) under the European Union’s Horizon 2020 research and innovation programme grant agreement ERC-2019-AdG 882679.
MSF is support by the Deutsche Forschungsgemeinschaft (DFG, German Research Foundation) under Germany's Excellence Strategy -- EXC-2094 ``Origins'' -- 390783311. GD acknowledges the funding by
the European Union - NextGenerationEU, in the framework of the HPC project – “National Centre for HPC, Big Data and Quantum Computing” (PNRR - M4C2 - I1.4 - CN00000013 – CUP J33C22001170001). 
We thank all participants of the Darkium SIDM \url{https://www.darkium.org/} Journal Club for helpful discussions.
The authors thank the organisers of the Pollica 2023 SIDM workshop for the interesting discussions.
AR and GC thank the support from INAF theory Grant 2022: Illuminating Dark Matter using Weak Lensing by Cluster Satellites, PI: Carlo Giocoli. 
\end{acknowledgements}

%
\bibliographystyle{aa} 
 \bibliography{bolo,globalbibs} 
%

\begin{appendix}

\section{Profile evolution}
\label{ap:evo}

 \begin{figure}
\includegraphics[width=1.1\linewidth]{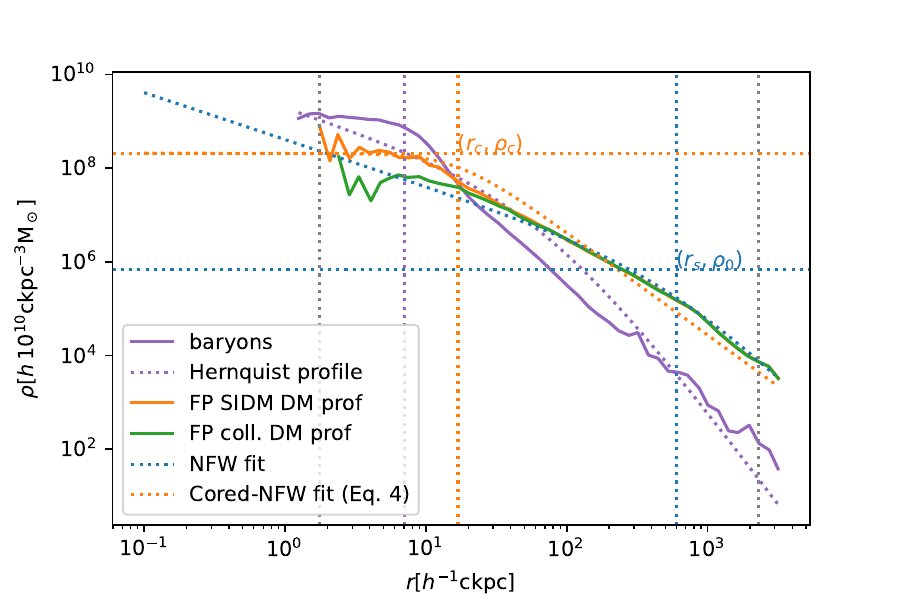}
\caption[h]{Density profile of D16 FP galaxy cluster at $z=0.2.$ We show its dark matter density profile (orange solid line); the corresponding NFW fit (dashed blue line) together with the values of $r_s$ and $\rho_s$ (vertical and horizontal dashed blue lines respectively); the corresponding cored-NFW functional form~\citep[dashed orange line, see our Eq.~\ref{eq:core} or][Eq. A5]{2021PhRvD.104j3031Yang}, together with the values of $\rho_c$ and $r_c$  (vertical and horizontal dashed orange lines respectively); the profile of baryons (solid purple line); the corresponding Hernquist profile (dotted purple line); and the DM profile of the corresponding DMO simulation (solid green line).  }
\label{fig:check}
\end{figure}

 \begin{figure*}
\includegraphics[width=0.246\linewidth]{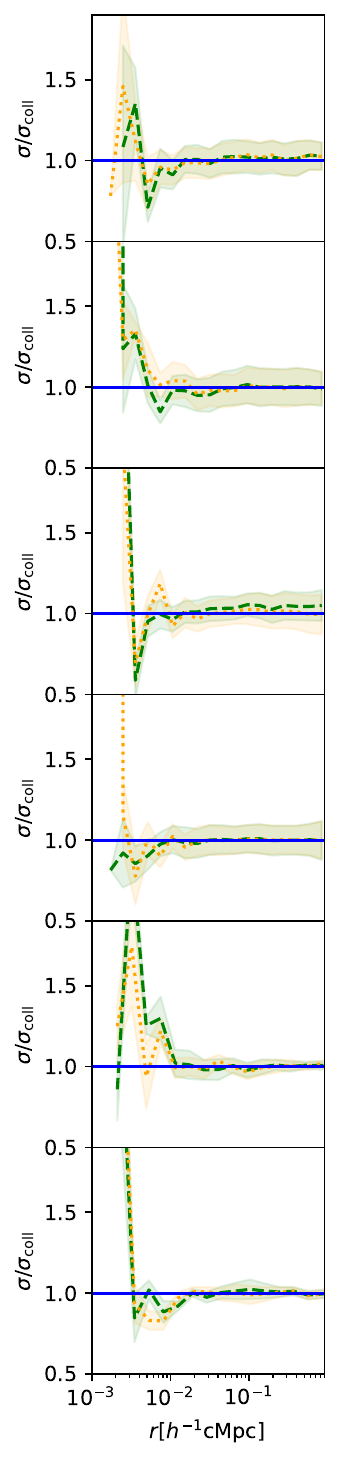}
\includegraphics[width=0.6\linewidth]{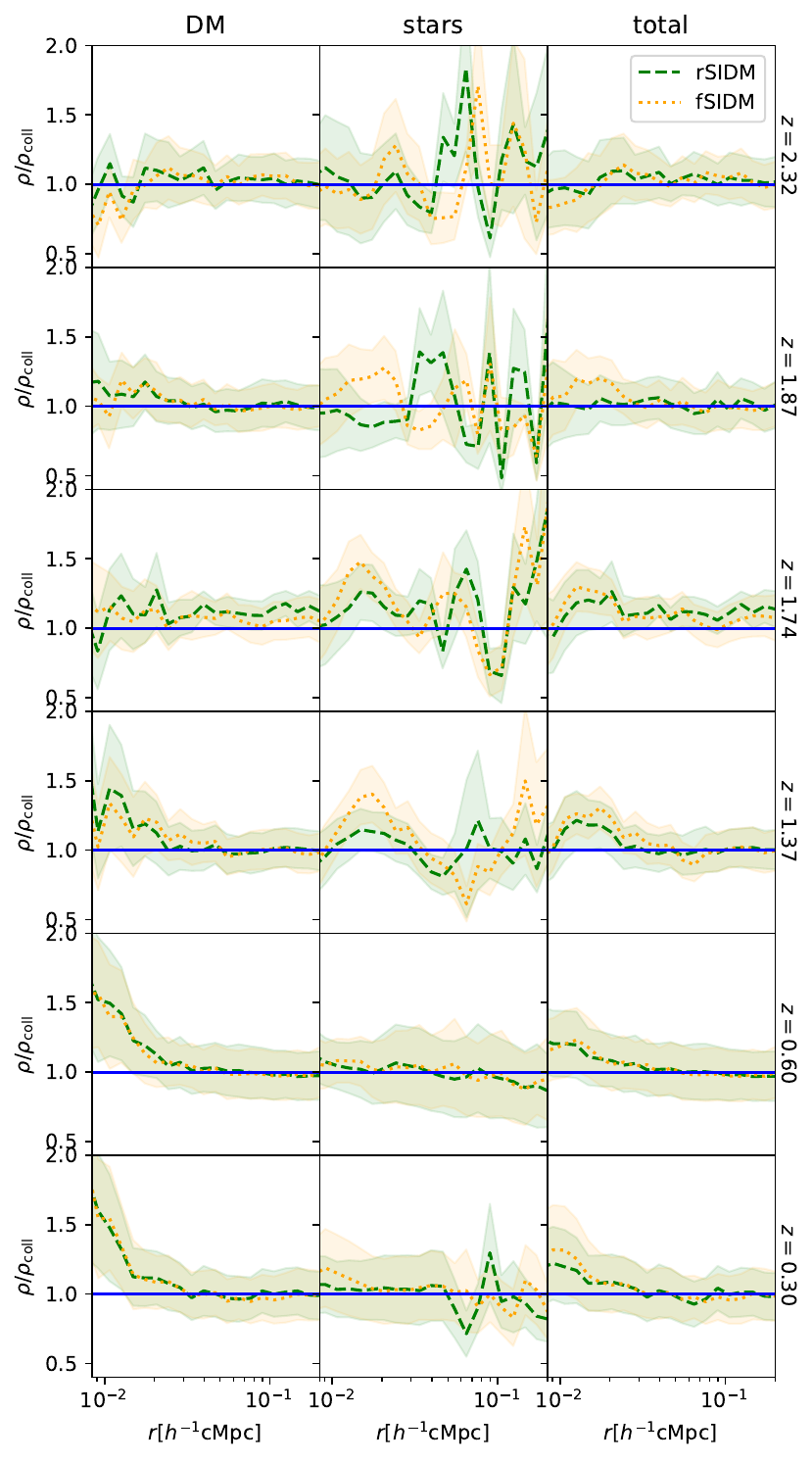}
\caption[h]{Relative velocity dispersion and density profiles of FP SIDM runs vs. FP collisionless DM runs. Each row represents a redshift slice. The left panel show the relative dispersion profile, while the right panels report the relative density profiles (of dark matter, stars, and total matter, respectively). Green dashed lines report values for rSIDM, while orange dotted lines report values for fSIDM. Shaded area represent the error on the mean of the $6$ regions.
Note that the velocity dispersion relative difference has a different radial range compared to the density relative difference (see discussion).}
\label{fig:evo}
\end{figure*}

In this Appendix we provide additional details on the central increase of the DM density and velocity dispersion.
First of all, we made sure that we have under control the computation of the central density $\hat\rho_c$ and of the central density of the baryons. 

In Fig.~\ref{fig:check}, we present the DM density profile of the FP run of the D16 galaxy cluster at the lowest redshift slice. There, we can see that the DM has a core and that the core is well fitted by the profile in Eq.~\ref{eq:core} (notice how the vertical and horizontal orange dashed lines capture the core density and radius of the orange solid line).
We also show that our NFW fits are properly capturing the DM profile (note how the blue dashed lines capture the orange solid line down to a few tens of kpc).
Finally, we show that the core of baryons is well captured by a radius of $\approx h^{-1}10{\rm ckpc}$ as stated in Sect~\ref{sec:cc}.

We now provide more details on the time evolution of the central increase of DM density and circular velocity.
The motivation behind this extended study is that in Section \ref{sec:cc}, we found that SIDM profiles on FP clusters have a higher central density compared to collisionless DM profiles of FP clusters. Therefore, we believe it is interesting to show the time evolution of both the velocity dispersion (Figure~\ref{fig:evo} left panels) and density profiles (Figure~\ref{fig:evo} right panels, for dark matter, stars and total matter, respectively) profile ratio of SIDM FP runs vs. the collisionless FP run values stacked for our six galaxy clusters. 
We see that at high redshift ($z=2.3$), both SIDM and collisionless runs of FP clusters match both in terms of velocity dispersion and density profiles; therefore, SIDM and CDM produce similar high-redshift galaxy clusters.
As time passes, both the central density and the central velocity dispersion of SIDM runs increase (compared to collisionless FP runs), as well as the relative central velocity dispersion.

Note that the velocity dispersion relative difference has a radial range that starts at ${\rm ckpc}$ scale (see left column), while  the radial range of the relative density starts at $\approx10{\rm ckpc}$ (see right columns). We use two different radial range in order to better show the deviations of SIDM with respect to CDM. In fact, SIDM velocity dispersion deviates at  sub-softening scales (see, e.g., Fig.~\ref{fig:profivel}), while density profiles deviate at much larger radii.

\end{appendix}
\end{document}